\documentclass[10pt,letterpaper,twocolumn]{article} 

\usepackage{ol2}
\usepackage[draft,implicit=false]{hyperref}
\usepackage{amsmath}
\usepackage{amssymb}
\usepackage{color}

\begin{document}

\twocolumn[ 

\title{Spontaneous creation and annihilation of temporal cavity solitons in a coherently-driven passive fiber resonator}


\author{Kathy Luo,$^{1}$ Jae K. Jang,$^{1}$  St\'ephane Coen,$^{1}$ Stuart G. Murdoch,$^{1}$ and Miro Erkintalo$^{1,2,*}$ }

\address{
$^1$Physics Department, The University of Auckland, Private Bag 92019, Auckland 1142, New Zealand \\
$^2$Department of Electrical Engineering, Princeton University, Princeton, New Jersey 08544, USA \\
$^*$Corresponding author: m.erkintalo@auckland.ac.nz
}

\begin{abstract}
We report on the experimental observation of spontaneous creation and annihilation of temporal cavity solitons (CSs) in a coherently-driven, macroscopic optical fiber resonator. Specifically, we show that CSs are spontaneously created when the frequency of the cavity driving field is tuned across a resonance, and that they can individually disappear at different stages of the scan. In contrast to previous experiments in monolithic microresonators, we are able to identify these dynamics in real time, thanks to the macroscopic dimensions of our resonator. Our experimental observations are in excellent agreement with numerical simulations. We also discuss the mechanisms responsible for the one-by-one disappearance of CSs.
\end{abstract}


 ] 

\noindent \looseness=-1 Temporal cavity solitons (CSs) are self-localized pulses of light recirculating in a continuously-driven passive nonlinear cavity~\cite{wabnitz_suppression_1993}. They are, in many respects, similar to their extensively investigated spatial counterparts~\cite{barland_cavity_2002,ackemann_fundamentals_2009}: beams of light persisting in nonlinear diffractive cavities. In particular, multiple independent CSs can simultaneously circulate in the resonator with arbitrary separations.

\looseness = -1 Temporal CSs were first observed experimentally by Leo~\emph{et al.}~\cite{leo_temporal_2010} in a continuously-driven loop of passive single-mode optical fiber. Subsequent studies have employed similar configurations to examine in detail their characteristics and dynamics~\cite{jang_ultraweak_2013, leo_dynamics_2013,jang_observation_2014}, and the ability to manipulate them has also been demonstrated~\cite{jang_temporal_2014}. Temporal CSs have further attracted interest in the context of monolithic \emph{microresonators}. In particular, it was proposed that ``Kerr'' frequency combs generated in such devices~\cite{delhaye_optical_2007, kippenberg_microresonator-based_2011} can be associated with temporal CSs~\cite{leo_temporal_2010, coen_modeling_2013, coen_universal_2013, chembo_spatiotemporal_2013}. Whilst signatures of phase-locked comb states had been observed in earlier microresonator experiments~\cite{herr_universal_2012, saha_modelocking_2013}, it is only recently that convincing proof of temporal CSs has been reported~\cite{herr_temporal_2014, brasch_photonic_2014}. 

\looseness=-1 Fiber cavities share many things in common with monolithic microresonators, and the nonlinear dynamics within both devices can be modelled using the same Lugiato-Lefever equation (LLE)~\cite{haelterman_dissipative_1992,matsko_mode-locked_2011, coen_modeling_2013, coen_universal_2013, chembo_spatiotemporal_2013}. Yet, the means by which temporal CSs have been created in these systems differ dramatically. In fiber resonators, CSs have been controllably excited by inducing a localized phase-shift on the resonator continuous-wave (cw) steady-state, either all-optically through cross-phase modulation~\cite{leo_temporal_2010,jang_ultraweak_2013,leo_dynamics_2013,jang_observation_2014,jang_temporal_2014}, or electronically by phase-modulating the cavity driving field~\cite{jang_writing_2015}. In contrast, in microresonator experiments, solitons have been produced by carefully tuning and then stabilizing an initially blue-detuned pump laser across a cavity resonance~\cite{herr_temporal_2014, brasch_photonic_2014}. Simulations and theoretical analyses reveal that in this case the solitons emerge spontaneously as the end-result of a complex sequence of bifurcations between different classes of solutions of the LLE~\cite{herr_temporal_2014,coen_universal_2013, lamont_route_2013}. Experimentally such soliton formation dynamics have only been inferred from indirect cavity transmission measurements~\cite{herr_temporal_2014, brasch_photonic_2014}: the large free-spectral ranges (FSRs) of microresonators prevent direct time-resolved measurements. Fiber cavities would not suffer from this issue, but similar spontaneous CS formation has not been experimentally observed in this context. This is somewhat surprising given that both systems obey the LLE, raising the question of whether spontaneous CS formation is unique to microcavities.

In this Letter we answer this question by reporting the experimental observation of the spontaneous creation and annihilation of temporal CSs in a coherently-driven passive fiber resonator. In contrast to experiments in microresonators, the macroscopic dimensions of our fiber cavity allows us to directly resolve the roundtrip-by-roundtrip soliton dynamics using a fast real-time oscilloscope. Significantly, we are able to directly capture the individual annihilation of temporal CSs at different stages of the scan. All our experiments are in very good agreement with numerical simulations.

\looseness=-1 We begin by briefly recalling the dynamics associated with the spontaneous formation of temporal CSs~\cite{herr_temporal_2014,coen_universal_2013}. To this end, we introduce the normalized parameters $X = P_\mathrm{in}\gamma L\theta\mathcal{F}^3/\pi^3$ and $\Delta = \mathcal{F}\cdot(2\pi m-\phi)/\pi$ describing, respectively, the power of the driving laser and its phase detuning from the closest cavity resonance (with order $m$)~\cite{leo_temporal_2010, coen_universal_2013}. Here $P_\mathrm{in}$ is the driving power, $\gamma$ the nonlinear coefficient, $L$ the cavity length, $\theta$ the input coupler power transmission coefficient, $\mathcal{F}$ the cavity finesse and ${\phi=\beta_0L}$ the linear phase accumulated by the intracavity field over one roundtrip, with $\beta_0$ the propagation constant at the pump frequency. Because of the Kerr nonlinearity, the peak of the cavity resonance is tilted from $\Delta = 0$ to $\Delta = X$~\cite{coen_universal_2013}. To reach the resonance, the driving laser approaches it from $\Delta < 0$. The intracavity field is cw until the detuning exceeds the threshold of modulation instability (MI), $\Delta_\mathrm{MI} = 1-(X-1)^{1/2}$. MI then leads to the break-up of the cw field into an ultrahigh repetition rate periodic pattern. This pattern may initially be stable, but typically destabilizes for increasing $\Delta$. When $\Delta$ is sufficiently large so that the cw cavity response is multivalued ($\Delta>\Delta_\mathrm{CS}$), some of the temporal pulses associated with MI reshape into CSs. Depending on the pump-cavity parameters~\cite{leo_dynamics_2013}, the CSs may exhibit unstable ``breathing'' behaviour for detunings close to $\Delta_\mathrm{CS}$, but eventually stabilize as $\Delta$ increases. Owing to their emerging from unstable MI, the initial number and configuration of solitons in the cavity is random. As the laser further approaches the tip of the tilted resonance, however, the solitons can disappear one at a time, giving rise to distinct ``steps'' in the cavity transmission~\cite{herr_temporal_2014,brasch_photonic_2014}. Finally, when $\Delta\gtrsim \pi^2X/8$, CSs cease to exist~\cite{barashenkov_existence_1996, herr_temporal_2014}, and the intracavity field returns to the homogeneous cw state.

\begin{figure}[t]
\centering
\includegraphics[width=\columnwidth,clip = true]{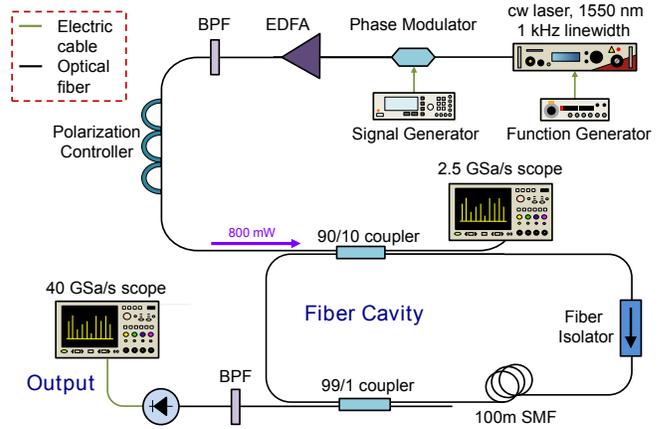}
\vskip -4pt
\caption{\small{(Color online) Schematic illustration of the experimental setup. }}
\label{setup}
\vskip -8pt
\end{figure}

\looseness=-1 To investigate these dynamics in a macroscopic fiber cavity, we use the experimental setup shown in Fig.~\ref{setup}. The basic configuration is similar to that used in~\cite{jang_ultraweak_2013,jang_temporal_2014}, and corresponds to a 100~m long (2~MHz FSR) single-mode fiber resonator with $\gamma = 1.2~\mathrm{W^{-1}km^{-1}}$, $\beta_2 = -21.4~\mathrm{ps^2km^{-1}}$ and a finesse of $\mathcal{F} \approx 20.3$. The resonator is driven with a narrow linewidth cw laser at 1550 nm, which is amplified to 800~mW with an erbium-doped fiber amplifier (EDFA). A 75~GHz full-width at half-maximum (FWHM) bandpass filter (BPF) is used to remove amplified spontaneous emission (ASE) noise before the laser is coupled into the resonator with a 90/10 fiber coupler. With these parameters, the normalised pump power $X \approx 2.6$. The laser frequency is tuned linearly such that a scan across the $\sim 100$~kHz-wide resonance occurs in about 10 ms (20000 roundtrips). During the scan, we monitor the average power transmitted by the resonator using a 300~MHz photodetector and a 2.5 GSa/s oscilloscope. In addition, one percent of the circulating field is extracted each roundtrip for monitoring with a 12.5~GHz AC-coupled photodetector connected to a faster 40 GSa/s oscilloscope. This allows the evolution of the temporal intensity profile to be examined. In this context, the 500~ns (i.e., $\mathrm{FSR}^{-1}$) roundtrip time of our cavity is so large that, unlike in microresonator experiments, the roundtrip-by-roundtrip evolution can easily be discerned.  Before detection, the extracted field passes through a narrow (0.6 nm FWHM) BPF  centered at 1551 nm. This removes the cw component at the driving wavelength of 1550 nm, thereby improving the signal-to-noise ratio of the measurement~\cite{leo_temporal_2010}.  In addition, the 1~nm offset from the driving wavelength ensures that all signals measured by the fast oscilloscope represent short temporal features.

As the laser is scanned over the resonance, the intracavity cw power reaches comparatively high levels, which can lead to significant stimulated Brillouin scattering (SBS) effects. Indeed, we find that an intracavity fiber isolator, typically used to suppress SBS in fiber CS experiments~\cite{leo_temporal_2010,jang_ultraweak_2013, jang_observation_2014}, is insufficient alone. We therefore additionally apply a 10~GHz sinusoidal phase modulation on the cavity driving field. This eliminates SBS, but also leads to the generated CSs being attracted to the peaks of the corresponding 10~GHz intracavity phase profile~\cite{jang_temporal_2014}.

\begin{figure}[b]
\centering
\includegraphics[width=0.95\columnwidth,clip = true]{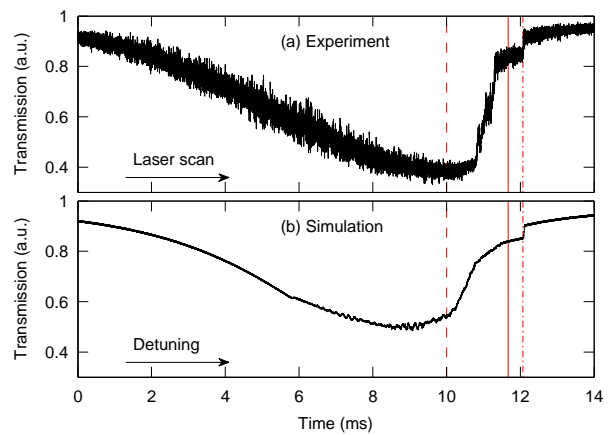}
\vskip -5pt
\caption{\small{(Color online) (a) Typical transmission measured when scanning the laser over a cavity resonance. (b) Corresponding results from numerical simulations. Vertical lines indicate approximate positions of real-time results in Fig.~\ref{real_time}.}}
\vskip -5pt
\label{resonance}
\end{figure}

Figure~\ref{resonance}(a) shows typical cavity transmission, measured at the reflection port of the 90/10 input coupler, as the laser frequency is scanned over the resonance. Corresponding results from LLE-based numerical simulations are shown in Fig.~\ref{resonance}(b). To enable fair comparison, the simulations use experimentally measured pump-resonator parameters and they take into account the pump phase modulation (10~GHz sinusoid with 0.3 rad amplitude) and the ASE noise on the driving laser (75~GHz bandwidth, 0.3~mW power). Good agreement is observed between the experiments and the simulations. In particular, in both cases the transmission decreases monotonously but then suddenly increases in two quasi-discrete steps, separated by a short-lived pedestal [highlighted by the solid vertical line in Fig.~\ref{resonance}].

\begin{figure}[t]
\centering
\includegraphics[width=0.95\columnwidth,clip = true]{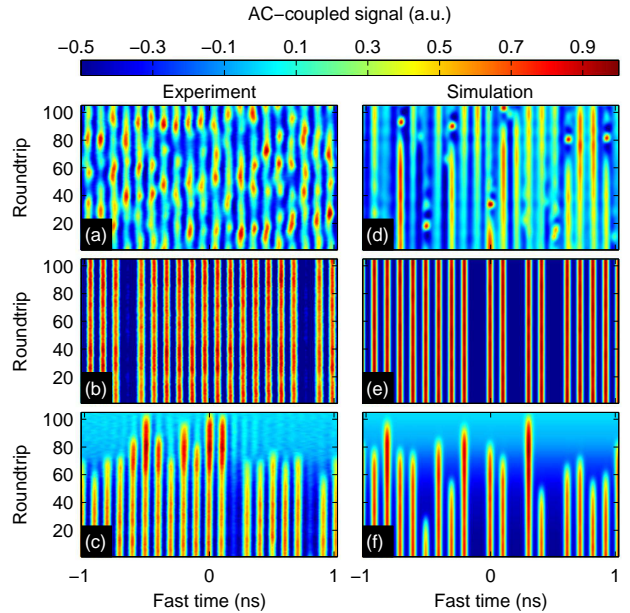}
\vskip -4pt
\caption{\small{(Color online) (a-c) Real-time roundtrip-by-roundtrip cavity dynamics measured at different points along the cavity resonance. (d-f) Corresponding results from numerical simulations, taking into account the detection optics and electronics.  The approximate positions of the 100-roundtrip snapshots are indicated as vertical lines in Fig.~\ref{resonance}: (a, d) dashed line; (b,e) solid line; (c, f) dash-dotted line.}}
\vskip -10pt
\label{real_time}
\end{figure}

To gain more insight, we have performed real-time measurements that resolve the roundtrip-by-roundtrip cavity dynamics. Unfortunately, the memory depth of our oscilloscope is limited to 2 million points at the highest sampling rate of 40 GSa/s, and so we can only perform measurements over 100 roundtrips during a single scan. Because the global dynamics occur over a much larger number of roundtrips (approximately 20000), we acquire several sets of data at carefully selected positions along the scan. This is achieved by triggering the fast oscilloscope from the resonator transmission measured by the slower oscilloscope. Typical experimental results at three different positions, approximately indicated with the vertical lines in Fig.~\ref{resonance}, are shown in Figs.~\ref{real_time}(a)-(c). Corresponding results from numerical simulations are shown in Figs.~\ref{real_time}(d)-(f). To facilitate comparison, the LLE-simulations are post-processed to take into account the offset filter as well as the limited 12.5 GHz bandwidth of the AC-coupled photodetector. Note also that for clarity the horizontal axes in Fig.~\ref{real_time} are limited to only show about 2 ns of the full 500 ns roundtrip.

Figures~\ref{real_time}(a, d) show dynamics close to the point of minimum transmission [dashed line at about 10.0~ms in Fig.~\ref{resonance}], and we clearly see features characteristic of the unstable MI regime: strong fluctuations from roundtrip to roundtrip. In stark contrast, a stable sequence of pulses is observed when the transmission lies within the pedestal between the two quasi-discrete steps [solid line at 11.7~ms in Fig.~\ref{resonance}], as shown in Figs.~\ref{real_time}(b, e). The pulses reside on a regular 10~GHz temporal grid defined by the pump phase modulation, yet do not form a periodic pulse train: clear holes --- unoccupied slots --- can be identified in the 10~GHz grid. These observations are indicative of temporal CSs. Indeed, the fact that the measurement is taken after the offset filter demonstrates the pulses to possess picosecond duration, as expected for CSs in this system~\cite{jang_ultraweak_2013}.  Moreover, the apparent holes prove that the field does not correspond to a periodic MI pattern. This is further highlighted by the dynamics recorded at the end of the scan, close to the second quasi-discrete step observed in the cavity transmission [dashed-dotted line in Fig.~\ref{resonance}]. Figure~\ref{real_time}(c) shows real-time dynamics in this region, and we see clearly how the different pulses die off at very different roundtrips. Results from numerical simulations are again in good agreement [Fig.~\ref{real_time}(f)]. Note that in our system the annihilation of individual CSs does not give rise to pronounced steps in the cavity transmission, as in~\cite{herr_temporal_2014, brasch_photonic_2014}. This is because there are about 5000 solitons in our cavity: annihilation of a single one has negligible effect on the average power.

\begin{figure}[t]
\centering
\includegraphics[width=0.95\columnwidth,clip = true]{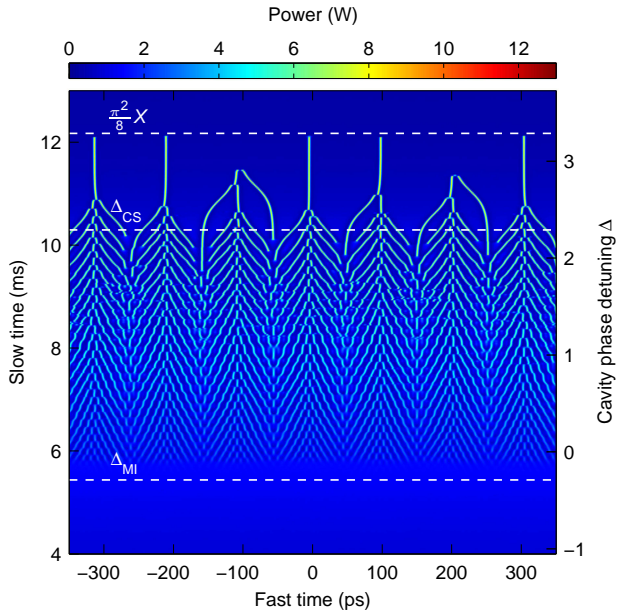}
\vskip -4pt
\caption{\small{(Color online) Numerically simulated intracavity dynamics without taking into account the off-set filter or the detection electronics. Dashed lines indicate, respectively, the limit detunings above which (i) MI can take place ($\Delta_\mathrm{MI}$), (ii) CSs can exist ($\Delta_\mathrm{CS}$) and (iii) CSs cease to exist ($\Delta = \pi^2 X/8$).}}
\vskip -11pt
\label{evolution_sim}
\end{figure}

The limited detection bandwidth, compounded by the need to suppress SBS with phase-modulation, hinders the direct interpretation of the measurements. To facilitate the analysis, we examine results from our numerical simulations in more detail. Figure~\ref{evolution_sim} shows the simulated intracavity evolution over a much larger number of roundtrips (and over a narrower time range) when neglecting the band-pass filter and the limited detection bandwidth. Here we also show, as dashed horizontal lines, the limit detunings $\Delta_\mathrm{MI} = 1-(X-1)^{1/2} = -0.26$ above which MI manifests itself, $\Delta_\mathrm{CS} = 2.29$ above which CSs can exist (found numerically) and $\Delta = \pi^2X/8=3.2$ above which CSs cease to exit. Note that  breathing CSs do not exist for our parameters.

As expected, when the detuning exceeds the MI threshold the intracavity cw field breaks-up into a periodic pattern. However, owing to the 10~GHz phase modulation, sub-pulses of the pattern continuously drift towards the peaks of the intracavity phase profile~\cite{jang_temporal_2014}. Pulses meeting at the peak merge into one, but new sub-pulses are continuously created at the phase minima to satisfy the periodic nature of MI. However, as soon as the detuning is sufficiently large for the pattern to reshape into isolated CSs ($\Delta > \Delta_\mathrm{CS})$, the creation of new pulses ceases. The remaining solitons drift to the phase peaks where, upon colliding with one another, they either merge into one~\cite{mcintyre_all-optical_2010} or annihilate. In fact, it is precisely these annihilations that give rise to, and explain, the gaps observed in the real-time measurements [Fig.~\ref{real_time}(b)]. Finally, we emphasize that these simulations clearly show that the pulses in Fig.~\ref{real_time}(e, f) correspond to picosecond temporal CSs, strongly suggesting the same for the corresponding experimental observations [Fig.~\ref{real_time}(b, c)].

We finally discuss the disappearing of the CSs at the end of the scan. In the simple LLE model all CSs are identical and they reside in identical environments~\cite{leo_temporal_2010}. All of them could therefore be expected to cease existing at the same detuning, in contrast to observations here and in microresonator experiments~\cite{herr_temporal_2014,brasch_photonic_2014}. The large observed spread alludes to the presence of a mechanism which changes the environment experienced by individual CSs, and therefore also their range of existence.  Depending on the pump-resonator configuration such a mechanism could be provided by interactions with another soliton, be it simple close proximity, or overlap with extended features arising, for example, from dispersive wave emission ~\cite{coen_modeling_2013, milian_soliton_2014}, or from perturbations induced by acoustic~\cite{jang_ultraweak_2013}, thermal or free-carrier effects~\cite{griffith_silicon-chip_2014, hansson_mid-infrared_2014}. We believe that, in our configuration, the necessary changes in local CS environment arise due to ASE noise on the cavity driving laser. Although we use a narrow (75~GHz) bandpass filter to remove ASE before coupling the field into the cavity, the residual noise nevertheless manifests itself as fluctuations with $\sim 20$~ps timescales. Given that our solitons are spaced by 100~ps, such fluctuations can readily give rise to locally varying environments. Indeed, if we neglect ASE noise in our simulations, we observe that all of the solitons decay at the same roundtrip.

To conclude, we have observed experimental signatures of spontaneous creation and annihilation of temporal CSs in a coherently-driven optical fiber resonator.  Thanks to the macroscopic dimensions of our cavity, we have been able to record the roundtrip-by-roundtrip soliton dynamics in real time, observing very good agreement with numerical simulations. Our results show that the spontaneous formation of temporal CSs is not unique to microscopic resonators, further solidifying the link between optical fiber cavities and microresonators.

\looseness=-1 We acknowledge support from the Marsden Fund of The Royal Society of New Zealand. M. Erkintalo also acknowledges support from the Finnish Cultural Foundation.

\end{document}